Is the ambient transverse velocity in a boundary layer flow non-zero or zero?

Asterios Pantokratoras
Professor of Fluid Mechanics
School of Engineering, Democritus University of Thrace,
67100 Xanthi – Greece
e-mail:apantokr@civil.duth.gr

**Abstract**
The concept of boundary layer flow, introduced in 1904 by Prandtl, is a popular field in Fluid Mechanics for engineers, physicists and mathematicians. In the classical Blasius boundary layer flow the transverse (normal) velocity reaches a finite value and remains constant in the free stream. If this finite vertical speed will be added to the free stream speed the resulting velocity in the free stream will be greater than the original velocity upstream of the plate which is irrational. This phenomenon has been marked and discussed in many Fluid Mechanics textbooks without a satisfactory explanation. In the present work we present a definite explanation by solving the complete Navier-Stokes equations. It is found that the real ambient transverse velocity is zero and not finite as it is predicted by the boundary layer theory. The same is valid in the classical free convection flow along a vertical isothermal plate.

1. **Introduction**

The most important development in Fluid Mechanics during the 20$^{th}$ century was the concept of boundary layer flow introduced by Prandtl in 1904. A boundary layer is that layer of fluid which forms in the vicinity of a surface bounding the fluid. Every time a fluid moves along a surface we have a boundary layer near the surface. Therefore boundary layers exist in the interior of water pipes, in sewer pipes, in irrigation channels, near the earth's surface and around buildings due to winds, near aeroplane wings, around a moving car, at the river bottom, inside the blood vessels and so on.

In the classical Blasius (1908) boundary layer flow there is no need for the ambient transverse velocity to be determined. Whatever value the solution produces for $v(x, y \to \infty)$ must be accepted. This transverse (vertical) velocity reaches a maximum in the free stream and not a zero value. Its value is

$$v_\infty = 0.860 u_\infty \sqrt{\frac{\upsilon}{x u_\infty}} \qquad (1)$$

The problem is well known in the literature (Arpaci and Larsen, 1984, page 221, Schlichting and Gersten 2003, page 158, Panton, 2005, page 502, White, 2006, page 236) and sometimes is named as the "Blasius paradox" (Lewins, 1999). Panton notes that " At first it appears unusual that $v$ does not go to zero as $y \to \infty$ " and devotes a paragraph to explain this phenomenon (Panton, 2005, page 521). In the present work we present a definite explanation to this phenomenon.

2. **Results and discussion**

We consider the flow along a horizontal, stationary, semi-infinite plate situated in a horizontal free stream with constant velocity. The Navier-Stokes equations of this flow are

continuity equation:  $\quad \dfrac{\partial u}{\partial x} + \dfrac{\partial v}{\partial y} = 0 \qquad (2)$

x-momentum equation:  $\quad u \dfrac{\partial u}{\partial x} + v \dfrac{\partial u}{\partial y} = -\dfrac{1}{\rho}\dfrac{\partial p}{\partial x} + \upsilon(\dfrac{\partial^2 u}{\partial x^2} + \dfrac{\partial^2 u}{\partial y^2}) \qquad (3)$

y-momentum equation:  $\quad u \dfrac{\partial v}{\partial x} + v \dfrac{\partial v}{\partial y} = -\dfrac{1}{\rho}\dfrac{\partial p}{\partial y} + \upsilon(\dfrac{\partial^2 v}{\partial x^2} + \dfrac{\partial^2 v}{\partial y^2}) \qquad (4)$

where $x$ is the horizontal coordinate, $y$ is the vertical coordinate, $u$ is the horizontal velocity, $v$ is the transverse (vertical) velocity, $p$ is the pressure, $\rho$ is the density and $\upsilon$ is the fluid kinematic viscosity. The boundary conditions are the following: The free stream boundary is located far away from the plate where

$$\frac{\partial u}{\partial y} = \frac{\partial v}{\partial y} = 0 \tag{5}$$

At the plate we have

$$u = 0, v = 0 \tag{6}$$

At the vertical surface passing through the plate leading edge we have

$$u = u_\infty, v = 0 \tag{7}$$

At the outlet surface of the calculation domain we have

$$\frac{\partial u}{\partial x} = \frac{\partial v}{\partial x} = 0, \ p = p_\infty \tag{8}$$

where $u_\infty$ is the free stream velocity and $p_\infty$ is the ambient fluid pressure.

We solved the equations (2)-(4) using the finite volume method of Patankar (1980). The SIMPLE (Semi-Implicit Method for Pressure-Linked Equations) was used. The algorithm may be summarized as follows:

- Set the boundary conditions.
- Compute the gradients of velocity and pressure.
- Solve the discretized momentum equation to compute the intermediate velocity field.
- Compute the uncorrected mass fluxes at faces.
- Solve the pressure correction equation to produce cell values of the pressure correction.
- Update the pressure field using an under-relaxation factor for pressure.
- Update the boundary pressure corrections.
- Correct the face mass fluxes.
- Correct the cell velocities.

The physical domain was discretized by using a uniform staggered grid with 200x100 nodes. The length domain L was twice of the width H (L=2H) in order to achieve parabolic conditions at the domain outlet. A hybrid scheme, based on the line-by-line method that combines the Thomas algorithm (TDMA) with the direct Gauss-Seidel method was used to solve the system of discretized equations. The under-relaxation factors were 0.3 for pressure-correction equation and 0.7 for the momentum equations. A double precision accuracy was used. The results are grid independent. The method is well known and has been used extensively in the literature with the commercial name FLUENT. Therefore no further information will be given here.

Figure 1 shows the variation of the longitudinal dimensionless velocity profiles $u/u_\infty$ as a function of the local Reynolds number $\text{Re}_x = \frac{u_\infty x}{\upsilon}$. The transverse coordinate $\eta = y\sqrt{\frac{u_\infty}{2\upsilon x}}$ is that used in the boundary layer theory (White, 2006, page 231). We see that as the Reynolds number increases the velocity profiles approach the Blasius profile and at $\text{Re}_x = 10^6$ the numerically calculated velocity profile is very close to the Blasius solution (theoretically the complete coincidence happens at infinite Reynolds number, see Schlichting and Gersten 2003, page 165). The approach of the Blasius profile is also a proof that our numerical solution procedure is satisfactory. In Figure 2 the variation of the dimensionless transverse velocity $v/u_\infty\sqrt{\text{Re}_x}$ is illustrated for different values of the Reynolds number. It is clearly seen that the transverse velocity reaches a maximum and then goes to zero away from the plate. The maximum increases gradually and finally becomes equal to the maximum predicted by the Blasius solution (0.8604, Panton, 2005, page 502) at $\text{Re}_x = 10^6$. In figure 3 we show our complete transverse velocity profile at $\text{Re}_x = 10^6$ and the corresponding profile of the Blasius solution. We see that the real transverse velocity becomes zero far away from the plate whereas the transverse velocity predicted by the boundary layer theory reaches a maximum and remains constant in the boundary layer.

Using the above mentioned code we solved the full Navier-Stokes equations of another classical boundary layer flow, i.e., the free convection along a vertical isothermal plate (Ostrach 1953). Figure 4 shows the streamwise dimensionless velocity ($\frac{uxGr^{-1/2}}{2\upsilon}$) profiles as a function of the transverse coordinate ($\eta = \left(\frac{Gr}{4}\right)^{1/4}\frac{y}{x}$) for Grashof number ($Gr = \frac{g\beta\Delta T x^3}{\upsilon^2}$) equal to $10^8$ and Prandtl number ($\Pr = \frac{\upsilon}{a}$) equal to 1. We see that the two profiles are very close. In figure 5 we show the dimensionless transverse velocity ($\frac{vx(Gr/4)^{-1/4}}{\upsilon}$) profile at $Gr = 10^8$ and the corresponding profile of the boundary layer theory (Ostrach 1953). We see that the real transverse velocity becomes zero far away from the plate whereas the transverse velocity predicted by the boundary layer theory reaches a minimum and remains constant in the boundary layer.

The boundary layer theory is an approximation of the real flow along a flat plate which is expressed correctly by the full Navier-Stokes equations. For that reason it is unable to express all the characteristics of the real flow.

3. **Conclusions**

The main result of the present work is that the ambient transverse velocity predicted by the boundary layer theory is not the real one. In reality the transverse velocity continues to change beyond the constant value, predicted by the boundary layer theory, and becomes zero away from the plate.

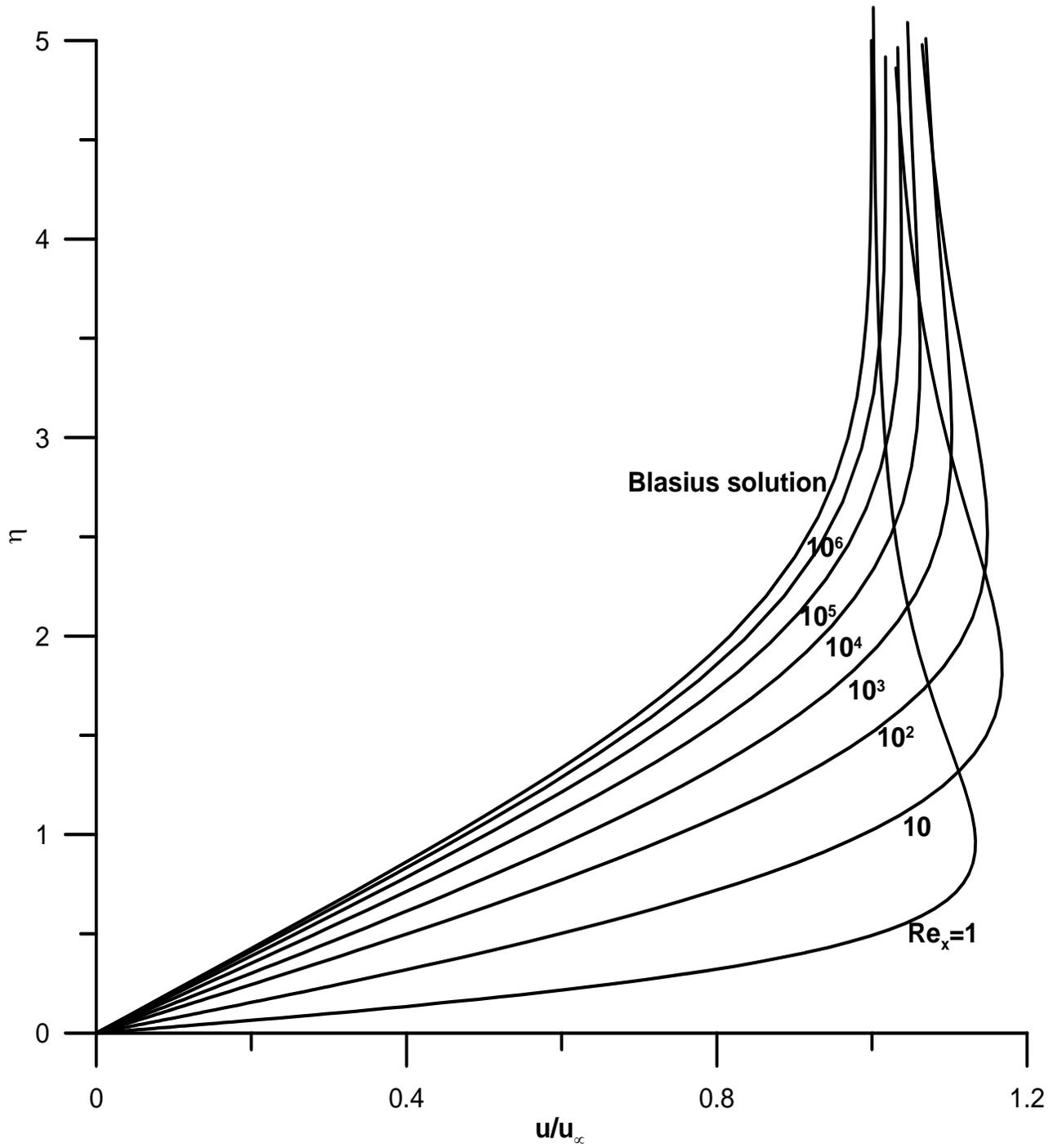

Figure 1. Variation of the longitudinal dimensionless velocity along the plate as a function of the local Reynolds number.

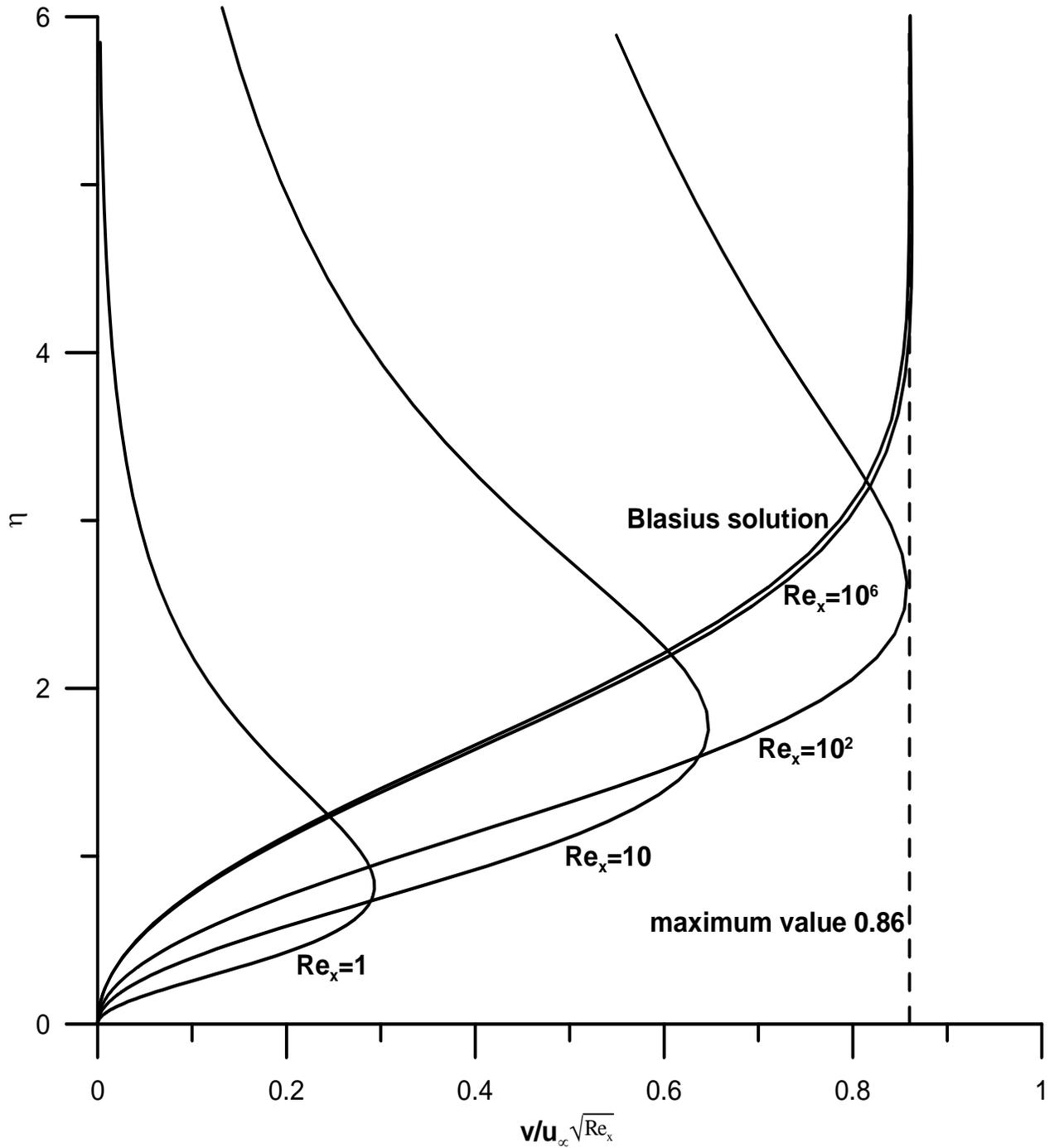

Figure 2. Variation of the transverse dimensionless velocity along the plate as a function of the local Reynolds number.

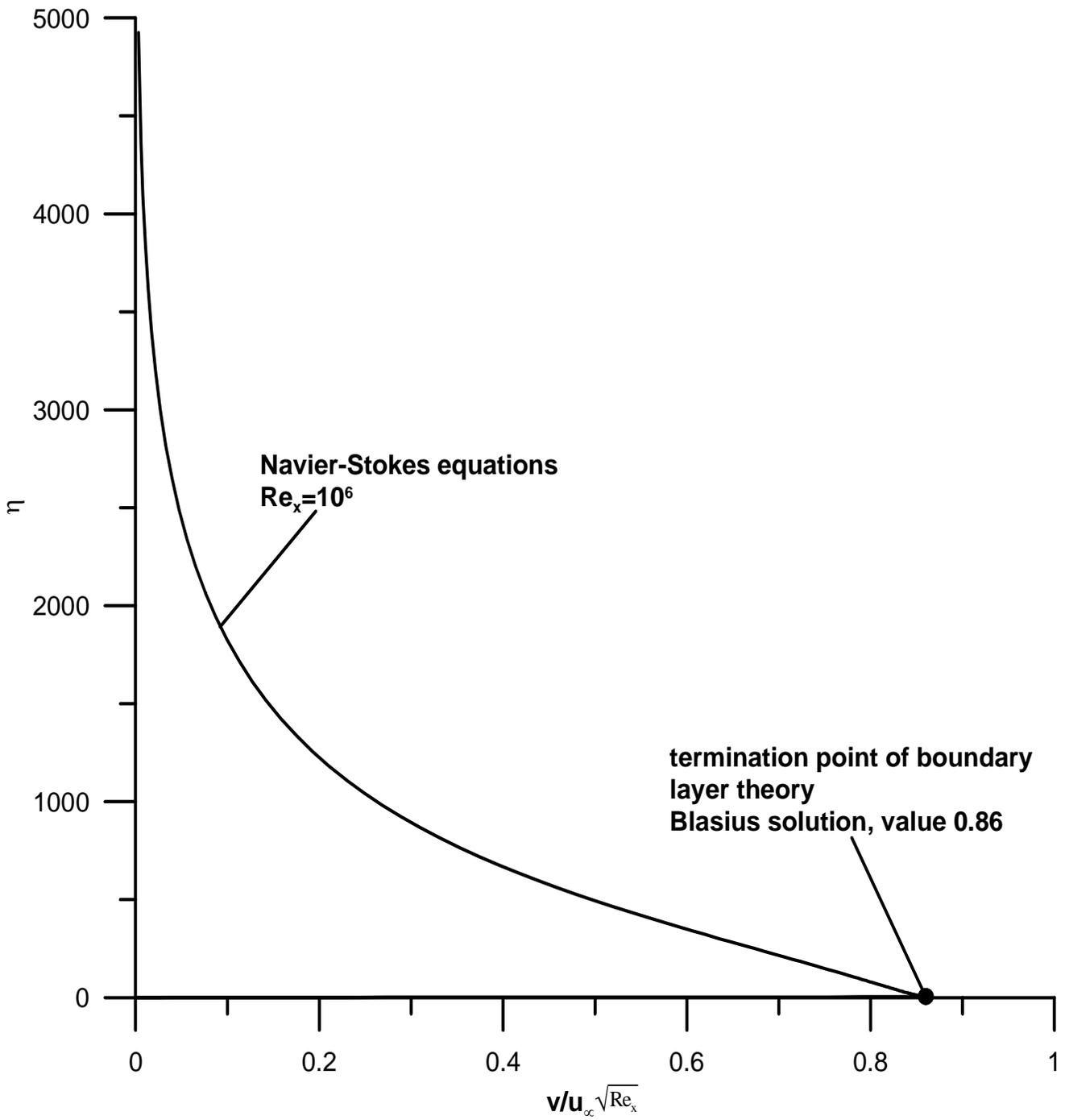

Figure 3. Comparison of the transverse dimensionless velocity profile of the Blasius solution with that of the Navier-Stokes equations at $Re_x = 10^6$.

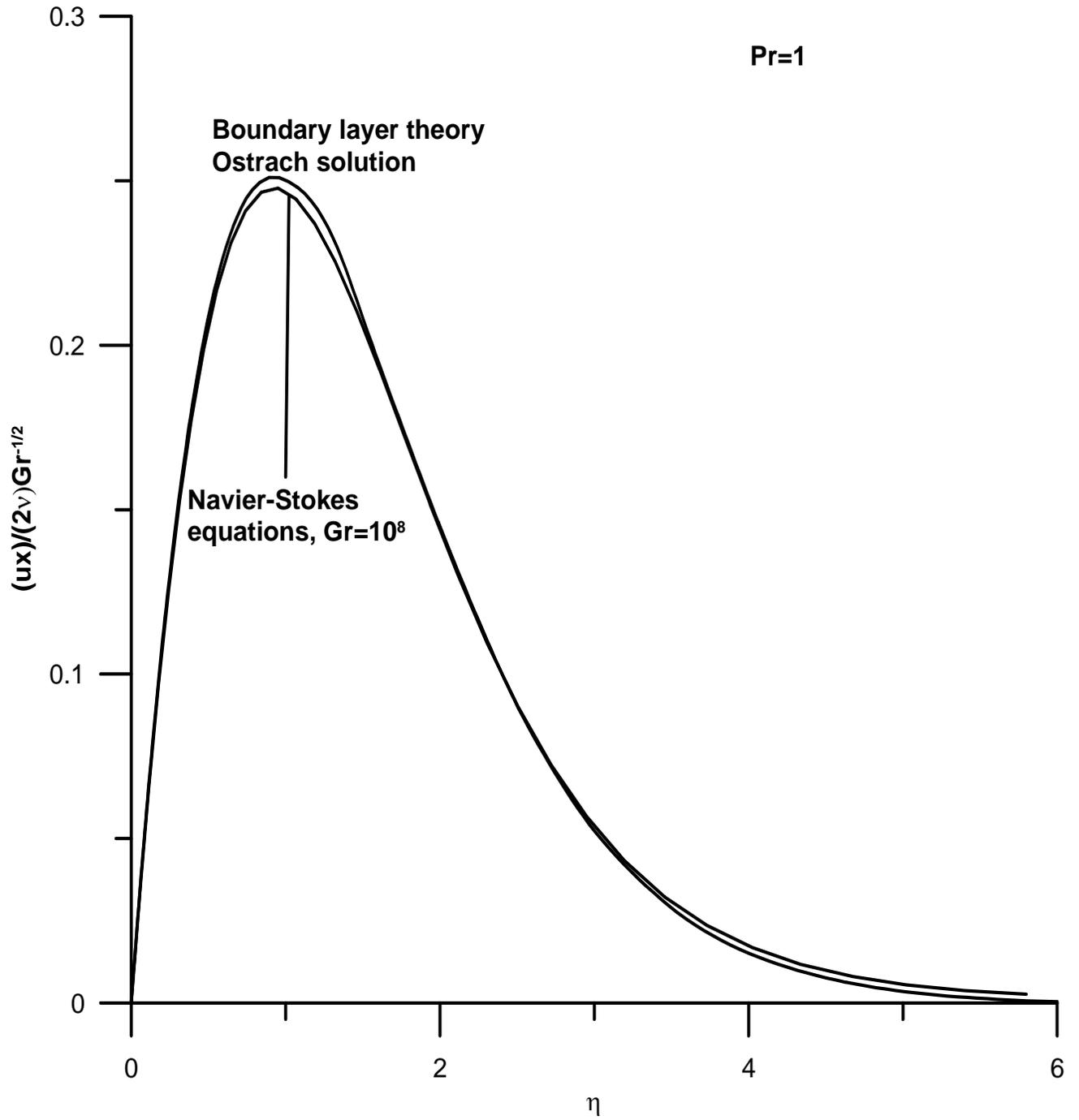

Figure 4. Comparison of the longitudinal dimensionless velocity profile of the Ostrach solution with that of the Navier-Stokes equations at $Gr = 10^8$.

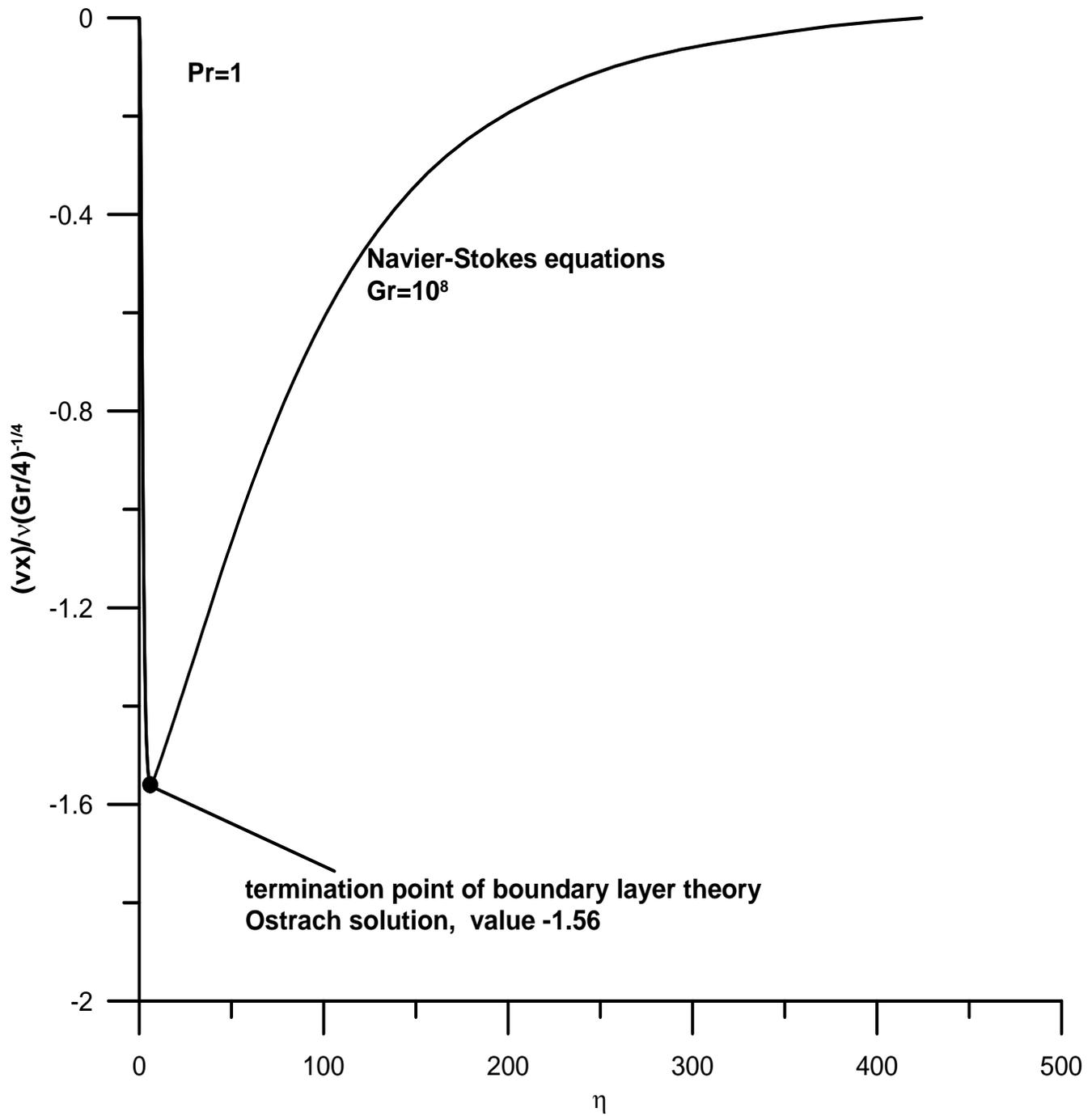

Figure 5. Comparison of the transverse dimensionless velocity profile of the Ostrach solution with that of the Navier-Stokes equations at $Gr = 10^8$.